\documentclass[11pt]{article}
\pdfoutput=1
\usepackage[utf8]{inputenc}
\usepackage{multirow}
\usepackage{amsmath, amsfonts, amssymb}
\usepackage{array}
    \newcolumntype{C}[1]{>{\centering\arraybackslash}m{#1}}
\usepackage{comment}
\usepackage{graphicx}
\usepackage{pdflscape}
\usepackage{psfrag}
\usepackage{amsthm}
\usepackage{enumerate}
\usepackage{arydshln}
\usepackage{pifont}
    
\usepackage{soul}
\usepackage{slashed}
\usepackage{subcaption}
\usepackage{mathrsfs}
\usepackage{ytableau}
 \usepackage{a4wide}
  \usepackage{tikz}
  \usepackage{tikz-cd}
  \usetikzlibrary{shapes.geometric}
  \usepackage{tcolorbox}
  \definecolor{dark-gray}{gray}{0.20}
  \definecolor{gray}{gray}{0.30}
  \definecolor{light-gray}{gray}{0.80}
  \definecolor{dark-red}{rgb}{0.7,0,0}
  \definecolor{dark-green}{rgb}{0.1,0.4,0}
  \definecolor{dark-blue}{rgb}{0.3,0.3,0.7}
  \definecolor{light-blue}{rgb}{0.8,0.8,1}
      \definecolor{swamp}{RGB}{240, 199, 197}
 
  \usepackage{pifont}

\usepackage{newunicodechar}
\usepackage{setspace}
 
\usepackage{ifthen}

\usepackage{longtable}

\newcommand{\be}{\begin{equation}}
\newcommand{\ee}{\end{equation}}
\newcommand{\eq}[1]{(\ref{#1})}

\def\be{\begin{equation}}
\def\ee{\end{equation}}
\def\bea{\begin{eqnarray}}
\def\eea{\end{eqnarray}}

\AtBeginDocument{

}

\numberwithin{equation}{section}

\usepackage{jheppub}
 
\usepackage{cleveref}
 
\hypersetup{
	colorlinks=true,
	linkcolor=dark-blue,
	citecolor=dark-red,
	urlcolor=dark-green,
	linktoc=page
}

\theoremstyle{definition}

\theoremstyle{remark}

\crefname{appendix}{Appendix}{Appendices}
 
\title{Neutrinos, B-L Symmetry and the Dark Dimension}
 
\author{Miguel Montero$^1$,}
\author{Cumrun Vafa$^2$}
\author{and Irene Valenzuela$^{1,3}$}
\affiliation{$^1$Instituto de F\'{i}sica Te\'{o}rica IFT-UAM/CSIC,
C/ Nicol\'{a}s Cabrera 13-15, Campus de Cantoblanco, 28049 Madrid, Spain}
\affiliation{$^2$ Jefferson Physical Laboratory, Harvard University, Cambridge, MA 02138, USA}
\affiliation{$^3$ CERN, Theoretical Physics Department, 1211 Meyrin, Switzerland}
\emailAdd{miguel.montero@csic.es, irene.valenzuela@cern.ch, vafa@g.harvard.edu}
\abstract{We consider realizations of a gauged B-L symmetry in the context of the Dark Dimension scenario, where the SM lives on a codimension one brane in 5d spacetime. The B-L can naturally be a bulk gauge symmetery leading to a global symmetry on the SM brane, and have its gauge anomaly canceled by charged bulk modes.  This naturally leads to the existence of 3 right-handed neutrinos propagating in the dark dimension. Allowing for spontaneous breaking of B-L in the bulk results in a massive gauge field with a very weak coupling $g_{B-L}\sim 10^{-10}$; for instance, B-L breaking at the weak scale gives $m_{B-L}\sim 100$ GeV, which is allowed by current bounds, as is a wide range of other values of the B-L breaking scale. The bulk sterile neutrinos are expected to have a mass of the order of the B-L breaking scale. Once the full Kaluza-Klein tower of bulk neutrinos is taken into account, the active neutrino masses turn out to be insensitive to the sterile neutrino mass, and reproduce the original Dark Dimension estimate $m_\nu\sim y^2\langle H\rangle^2/M_5$, which leads to the observed neutrino mass scale for Yukawa couplings $y\sim 10^{-3}$.
}

\begin{document}
\emergencystretch 3em
\hypersetup{pageanchor=false}
\makeatletter
\let\old@fpheader\@fpheader
\preprint{IFT-25-158,
CERN-TH-2025-255}
 
\makeatother
 
\maketitle
 
\hypersetup{
    pdftitle={},
    pdfauthor={},
    pdfsubject={}
}

\newcommand{\remove}[1]{\textcolor{red}{\sout{#1}}}
 
\newcommand{\red}[1][\text{(check)}]{\textcolor{red}{#1}}
 
% Notation commands
\newcommand{\D}[1][\gamma]{\mathbf{D}_{\bf #1}}
\newcommand{\bvec}[1][\gamma]{\vec{b}_{\bf #1}}
\newcommand{\RFM}[2][T]{%
  \ifthenelse{\equal{#1}{T}}%
    {\frac{T^{#2}}{\Gamma}}%
    {\frac{\mathbb{R}^{#2}}{\mathcal{B}}}%
}
\newcommand{\Tr}[2]{{\rm Tr_{\bf #1}}(#2)}
\newcommand{\TrB}[1]{\Tr{B}{#1}}
\newcommand{\TrF}[1]{\Tr{F}{#1}}
\newcommand{\Vcas}{V_{\text{Cas}}}
 
\section{Introduction}
\label{sec:introductions}
 
The Dark Dimension scenario \cite{Montero:2022prj} has been motivated from Swampland principles, and in particular from the Distance Conjecture \cite{Ooguri:2006in,Lust:2019zwm} applied to Dark Energy. In a nutshell, it proposes the existence of one mesoscopic extra dimension in our universe with diameter in the $0.1-10\ \mu m$ range, which is significantly larger than other extra dimensions. This large dimension (dubbed the Dark Dimension) is motivated from the smallness of dark energy $\Lambda$ and leads to the existence of a light infinite tower of states whose mass is correlated with the value of the dark energy. Theoretical and experimental bounds single out this tower to be a Kaluza-Klein tower of mass $m_{\rm KK}\sim \Lambda^{1/4}$, signaling decompactification of only one extra dimension \cite{Montero:2022prj}\footnote{See \cite{Anchordoqui:2025nmb,Hardy:2025ajb} for a recent discussion on the borderline experimental feasability of having two large dark dimensions if the diameter is smaller $\lesssim 0.3 \ \mu m$.}. In this scenario, the SM fields should live in a brane localised in the Dark Dimension. Although it remains as an open question why the Dark Energy is so small in the first place, it provides a quantum gravity link between the Dark Energy and diverse aspects of Particle Physics.
 
This has led to a very rich phenomenology addressing open puzzles in Particle Physics and Cosmology. For instance, the Dark Dimension scenario has led to new models for dark matter and cosmology \cite{Anchordoqui:2022txe,Gonzalo:2022jac,Anchordoqui:2022tgp,Anchordoqui:2022svl,Anchordoqui:2023oqm,Noble:2023mfw,Anchordoqui:2023wkm,Anchordoqui:2023tln,Law-Smith:2023czn,Cui:2023wzo,Obied:2023clp,Anchordoqui:2024akj,Anchordoqui:2024dxu,Gendler:2024gdo,Anchordoqui:2024gfa,Heckman:2024trz,Basile:2024lcz,Anchordoqui:2024xvl,Anchordoqui:2025xug,Antoniadis:2025rck,Eller:2025lsh}. In particular \cite{Bedroya:2025fwh}, based on the earlier model \cite{Agrawal:2019dlm}, leads to the best known fit of a physical model to the recent cosmological observations of evolving dark sector \cite{DESI:2025zgx,DESI:2025fii}.  The Dark Dimension scenario also has a distinct experimental prediction for deviations from Newton's inverse square law at micron length scales (see \cite{Adelberger:2003zx,ParticleDataGroup:2022pth,Lee:2020zjt} for reviews) which is currently being pursued by a number of groups.

It is thus natural to ask how various Beyond Standard Model (BSM) questions fit in this scenario.  For example, see \cite{Heckman:2024trz} for a discussion of how GUTs may work with this scenario. Another glaring question which demands a BSM explanation is the existence of mass for neutrinos.  A proposal of how neutrinos may fit was already presented in \cite{Montero:2022prj}, along the lines of similar discussions in Large Extra Dimension scenarios, see \cite{Arkani-Hamed:1998wuz,Dienes:1998sb}, by allowing massless bulk five-dimensional right-handed neutrinos (see \cite{Antoniadis:2025rck} for a recent study suggesting these may be within experimental reach).  As we will review below, this proposal has the benefit that it utilizes (but does not explain from fundamental principles) the observed coincidence between the mass scale for neutrinos and that of Dark Energy.  On the other hand, postulating that there are at least three massless fermions in the bulk just to play the role of right-handed neutrino and provide mass to the neutrinos seems somewhat ad hoc.
 
The main aim of this brief note is to explore alternatives or variants to this proposal which address in addition some other puzzles of the Standard Model.  In particular, why is an anomalous B-L gauge symmetry, a good global symmetry of the SM, and how does neutrino physics fit with this? Since there are no global symmetries in quantum gravity theories, it is natural to expect that there is a $U(1)_{B-L}$ gauge symmetry in nature (spontaneously broken or not), with some additional ingredients to cancel the gauge anomalies.  As we will see in this note, in the Dark Dimension scenario this can happen in one of two ways:  Anomaly inflow from the bulk, or some additional charged states in the SM brane (a possibility that naturally arises in GUT-like models). We explore the anomaly inflow option (which itself has two variants, depending on whether the bulk B-L gauge field is made massive via a Green-Schwarz or Higgs mechanism) and find that the B-L Higgs option naturally explains why there are only three massless neutrinos, yields a value for the gauge field coupling which is compatible with experimental bounds for a wide range of masses, \emph{and}, once the full tower of bulk neutrino modes is taken into account, leads to active neutrino masses which are insensitive to the mass of the bulk sterile neutrinos and coincide with the original Dark Dimension estimate. Thus, a bulk B-L gauge field, made massive by a Higgs mechanism, is a quite appealing possibility enhancing the predictive power of the Dark Dimension scenario.
 
The organization of this note is as follows:  In section \ref{s2} we review the option for giving mass to neutrinos by bulk right-handed neutrinos, including its strengths and weaknesses.  In section \ref{s3} we study the bulk B-L option and explain how it can lead to an appealing explanation for neutrino masses.  In section \ref{s4} we end with some concluding remarks.  Some technical discussions are presented in the appendix.

\section{Neutrino masses in the Dark Dimension}\label{s2}
We begin by reviewing the simplest ways to generate neutrino masses in the context of the Dark Dimension scenario. Of course, the simplest possibility is that the SM fields embed in a standard GUT scenario with a gauge group like $SO(10)$, where the right-handed neutrinos are unified with the rest of the SM fields. Such a model is completely compatible with the Dark Dimension despite the fact that GUT scale is far above the 5d Planck scale \cite{Heckman:2024trz}, as the heavy GUT states are realized as stretched strings.  But here, we would like to explore also other options which may be even more natural in the context of the dark dimensions scenario.
 
In the Dark Dimension (DD) scenario, the SM fields live in a localized brane. After EW symmetry breaking, this contains massless left-handed neutrinos, for which masses must be generated. One simple possibility, and quite natural from an extra-dimensional point of view, arises by assuming the existence of massless, five-dimensional bulk fermions. In four-dimensional language, these generate a tower of sterile, right-handed neutrinos with mass scale that can be naturally of the order of tower scale $m_{\nu_s}\sim\Lambda^{1/4}$, which roughly coincides with the neutrino mass scale \cite{Montero:2022prj}.  The idea that bulk fermions can play such a role was also considered in the Large extra dimension models \cite{Dienes:1998sb,Arkani-Hamed:1998wuz}, though in that context one needs to assume unnaturally one dimension is much larger than the rest, whereas this is automatic for the Dark Dimension scenario. In the DD scenrio, one expects the light left-handed neutrinos on the GUT brane to couple to the KK tower of right-handed ones in the bulk (see e.g. \cite{Carena:2017qhd}). Schematically this leads, for each flavor, to an effective $2\times 2$ mass matrix of the form
\begin{equation}\mathcal{M}=\left(\begin{array}{cc} 0&\frac{y \langle H\rangle}{\sqrt{ M_5R}}\\
 \frac{y \langle H\rangle}{\sqrt{ M_5R}}&m_{\nu_s}
\end{array}\right),\label{matrix}\end{equation}
where $R\sim 1/m_{KK}$ is the size of the extra dimension, $M_5\sim \Lambda^{1/12}\sim  10^{10}$ GeV is the 5d Planck scale, $\langle H\rangle$ is the Higgs vev and $y$  the Yukawa coupling of the SM brane with the bulk states.  The $1/{\sqrt R}$ comes from the wave function normalization of the bulk 5d fermions, assuming a homogeneous space, and we have replaced the infinite tower of right-handed modes by a single one for simplicity, as it does not change the qualitative features. See \cite{Montero:2022prj,Dienes:1998sb} for details in the general case.
 
 Assuming $m_{s}$ is bigger than or equal to the off diagonal term we can estimate the lower eigenvalue corresponding to the active neutrino to be
\begin{equation} m_\nu\approx   \frac{y^2\langle H\rangle^2}{M_5},\end{equation}
modulo $\mathcal{O}(1)$ factors that are controlled by the effective number of states in the tower (a more precise treatment of the full tower, including a bulk mass for the neutrino, is given in Section \ref{good}).
This is an expression similar to the usual see-saw mechanism with a Yukawa coupling of order $y\sim 10^{-3}$. The rest of the mass eigenstates will have an increasing mass starting at $m_{KK}$. See \cite{Antoniadis:2025rck,Eller:2025lsh} for recent analysis on the experimental signatures of these bulk right-handed neutrinos in the context of the Dark Dimension scenario.
 
To obtain the correct value of the neutrino masses one needs an explanation of the weak scale from the Dark Dimension perspective. This was addressed in \cite{Gonzalo:2022jac}, by imposing that the moduli of the SM brane has had enough time during the cosmological evolution of the universe to decay away, which leads to their mass being $\gtrsim\Lambda^{1/6}$. If we take this as the ballpark for $H$, it leads to $m_\nu/y^2 \gtrsim \Lambda^{2/6}/\Lambda^{1/12}\sim \Lambda^{1/4}$, leading to a potential explanation of the neutrino mass.
Even though the parameter range works very well, modulo the somewhat small value for the Yukawa coupling $y$, it should be emphasized that the bound $m_{SM}\gtrsim \Lambda^{1/6}$ is not a built-in feature of the Dark Dimension scenario -- rather, it is imposed on the actual model in order to realize a sensible cosmology compatible with observations.  Therefore the observed coincidence $m_\nu\sim m_{KK}$ is anthropic in this model -- it is only there because, if it was not, many scalars would not have stabilized by now, perhaps preventing the emergence of observers such as ourselves. By contrast, the coincidence $m_{KK}\sim \Lambda^{1/4}$ is a fundamental feature of the Dark Dimension scenario, irrespectively of other coincidences.
 
An unappealing feature of this scenario is the fact that one has to assume the existence of extra massless or light fermions
in the bulk, which may seem a bit ad hoc.  In fact, one needs {\it at least three } bulk fermions to give all active neutrinos mass. To see this, note that, assuming a thin SM brane, the coupling of an SM brane fermion $\nu_L$ to a bulk fermion $\psi$ can only involve the value of $\psi$ at the location of the SM brane (or possibly including derivative couplings),
\begin{equation} \int_{\text{SM brane}} \nu_L \cdot \psi\Big|_{SM brane}.\end{equation}
This implies that each bulk fermion only yields effectively a single lower-dimensional mode that couples to $\nu_L$, so that we cannot generate masses for the three SM left-handed neutrinos if starting with only one bulk fermion. We need to have at least three bulk fermionic fields to do so.
In contrast, if we give the SM brane some size, then a massive mode of the bulk fermion which has wavelength as the size of the SM brane may give an extra mode.  But the mass of such a mode can be estimated to be roughly $1/L$, where $L$ is the SM brane size. At this scale, we also expect to find the KK copies of the rest of the SM fields.  The fact that these have not been observed by LHC means that this mass scale has to be at least 10 TeV (and is expected to be of that order in the GUT realization of dark dimensions \cite{Heckman:2024trz}). Such a large value for $m_{KK}$ would lead, due to \eq{matrix}, to a neutrino mass
\begin{equation} m_\nu\sim \frac{L}{R}\frac{y^2\langle H\rangle^2}{M_5},\end{equation}
so that the two light neutrinos coupled to these KK modes have masses too small by a factor of $L/R\approx 10^{-12}$ and are therefore ruled out.
Therefore, for this picture to work and give mass to all neutrinos, it is absolutely necessary to have three bulk fermions for no apparent reason other than to give mass to neutrinos, which is one major shortcoming.
 
Another shortcoming  is that, in general, it does not solve the gauge anomaly of B-L symmetry.  In this scenario, the B-L is a global symmetry of the SM which is completely accidental (there is, though, one way to gauge this symmetry while having bulk neutrinos which we will discuss in the next section). It is thus natural to explore other ways to realize neutrino masses which are compatible with solving the gauge anomaly for the B-L gauge symmetry. In the next section we turn to studying this question and show that one may address this via an anomaly inflow mechanism from the five-dimensional bulk.

\section{The bulk B-L scenario}\label{s3}
In the previous Section, we have argued that just adding three right-handed neutrinos to the SM is rather ad-hoc. It also ignores the B-L symmetry of the SM.  We will now explore variants of the Dark Dimension scenario where the fate of B-L is taken into account. As explained in the Introduction, one possibility is always having a GUT scenario, where B-L is a SM brane gauge field which unifies with the SM at high energies. While this is certainly a viable possibility, its phenomenology in the context of Dark Dimension is not different from that of more general GUT models. In this Section, we will focus on phenomenology more specific to the Dark Dimension, in scenarios where B-L is directly related to a five-dimensional gauge field $A_B$. Since the SM has an anomalous spectrum with respect to B-L, the bulk sector will be constrained via anomaly inflow.
 
In principle, we could take the Dark Dimension to be an interval $S^1/\mathbb{Z}_2$ (as in a Horava-Witten setup \cite{Horava:1995qa,Horava:1996ma} or an orbifold in string theory). This has been discussed in the context of Dark Dimension in  \cite{Schwarz:2024tet} (see also \cite{Reig:2025dpz}), though what we have to say in this paper is not restricted to this case. If we take the fifth dimension to be an interval, the question arises as to what are the boundary conditions of the 5d gauge field $A_B$ at the endpoints. While there are infinitely many possibilities \cite{Arbalestrier:2025jsg}, we will focus here on two simple ones:\begin{itemize}
    \item The longitudinal component has Dirichlet boundary conditions, while the normal one has Neumann. Under dimensional reduction to 4d, only the axionic component survives. It then will get eaten up by a B-L gauge field localised on the brane via a Green-Schwarz mechanism.
    \item The component of $A_B$ normal to the boundary has Dirichlet boundary conditions, while the longitudinal component satisfies Neumann boundary conditions. Under dimensional reduction, this yields a four dimensional B-L gauge field.
\end{itemize}
Both of these are realized in the case where the boundary is understood as an $\mathbb{R}/\mathbb{Z}_2$ orbifold, depending on whether the vector $A_B$ transforms as a pseudovector or as an ordinary vector respectively with respect to the $\mathbb{Z}_2$ parity symmetry used in the quotient.  We can think of the discussion below as corresponding to these two cases, although the boundary condition discussion applies more generally. As we will see the vector case is more appealing phenomenologically.
 
\subsection{The pseudovector case}
As we will now see, the pseudovector case (where the longitudinal component has Dirichlet boundary conditions) leads to no bulk neutrinos  -- instead, B-L anomaly cancellation happens via anomaly inflow from the bulk, very much like in an $E_8$ Horava-Witten wall in string theory \cite{Horava:1995qa,Horava:1996ma}.  The similitude arises from the fact that the three-form $C_3$ in M-theory transforms as a pseudo-three-form under parity. In the four-dimensional language, the anomaly inflow leads to a Green-Schwarz mechanism \cite{Green:2012pqa}.
 
Let us begin describing the details of the scenario. Instead of having just the Standard Model fields, we will include an additional $U(1)$ to encode B-L charges. For instance, inspired by GUT with gauge group
\begin{equation}G_{\text{Brane}}=\frac{SU(5)\times U(1)_X}{\mathbb{Z}_5},\end{equation}
 we can take a $U(1)_X$ which is a certain linear combination of B-L and hypercharge,
\begin{equation}X=-2Y+5(B-L).\end{equation}
 For instance, breaking $SO(10)\rightarrow U(5)$ naturally produces the boson $X$.
 
If we do not include right-handed neutrinos, the $U(1)_X$  is anomalous. This is equivalent to the statement that $B-L$ is anomalous without the addition of right-handed neutrinos. This is similar to the $E_8$ gauge fields of the Horava-Witten wall. Just like in that case, the anomalies can be cancelled by inflow from the bulk. Specifically, let us include a 5d gauge field $A_B$, with $F_B=dA_B$ and 5d kinetic term
\begin{equation}\frac{1}{g_5^2}\int  F_B\wedge *F_B\label{a0}\end{equation}
The gauge coupling $g_5^2$ has units of length, and so it is natural to write
\begin{equation} \frac{1}{g_5^2}=\frac{M_5}{\alpha}\label{norsd}\end{equation}
where $\alpha$ is dimensionless and controls the ratio with the five-dimensional Planck's mass $M_5$.
 
The bulk field can be dualized to a 2-form $B_B$, with $H_B=dB_B$ and kinetic term
\begin{equation}\frac{1}{\tilde{g}_5^2}\int H_B\wedge *H_B,\quad  \tilde{g}_5=\frac{2\pi}{g_5}= \frac{2\pi}{\sqrt{\alpha}} M_5^{1/2}.\label{b0}\end{equation}
 
Now, we will postulate that the SM brane includes a coupling
\begin{equation} p\int_{\text{Brane}} B_B\wedge F_X\end{equation}
between the bulk and boundary fields. The equation of motion for $B_B$ then becomes
\begin{equation} d*H_B=dF_B= F_X\delta_{\text{Brane}},\end{equation}
which is solved if we impose the boundary condition
\begin{equation}F_B\vert_{\text{Brane}}=F_X.\label{bc0}\end{equation}
This is analogous to the Horava-Witten boundary condition $G_4\vert_{HW}=c_2^{E_8}$, and it allows us to cancel anomalies via inflow. In particular, if the bulk action for $A_B$ carries Chern-Simons terms,
\begin{equation} \int\frac{\mathcal{A}_1}{6} F_B\wedge F_B \wedge A_B -\frac{\mathcal{A}_2}{24}p_1(R)\wedge A_B,\label{chb}\end{equation}
where $\mathcal{A}_1,\mathcal{A}_2\in\mathbb{Z}$, then this cancels the boundary anomaly of the $U(1)_X$ gauge boson if the 6d polynomial associated to \eq{chb} is exactly the anomaly polynomial of $U(1)_X$.
 
Notice that what we have here is slightly different from standard anomaly inflow of the kind we have e.g. on heterotic strings. There, an anomaly of a global worldsheet symmetry coupled to a bulk gauge field is cancelled by a bulk Chern-Simons coupling. Here, we use the bulk to cancel the anomaly of a \emph{worldvolume} gauge field, that mixes with the bulk. One could say that $A_X$ is just the restriction of $A_B$ to the brane, but they don't have to agree, just their fieldstrengths have as in \eq{bc0}. In the HW case they are clearly different fields, as an $E_8$ gauge field cannot be regarded as a restriction of $G_4$ to a boundary (only at the topological level). In modern language, one identifies boundary conditions for the $U(1)$ gauge field $A_B$ with quantum field theories living in the boundary and equipped with a $U(1)_B$ global symmetry \cite{Arbalestrier:2025jsg}; in this case, the Chern-Simons terms of the bulk theory specify the mixed gravitational-$U(1)_B$ anomalies of the boundary theory.
 
We will now describe how the above manifests itself in the four-dimensional gauge theory. The bulk gauge field $A_B$ decomposes into a 4d gauge field $A_B^{(4)}$ and a $2\pi$-periodic axion $\phi_B$, as
\begin{equation} A_B=A_B^{(4)}+ \frac{dx}{R} \phi_B.\end{equation}
On an interval compactification, $A_B^{(4)}$ is projected out, since the existence of Chern-Simons terms \eq{chb} force again parity transformations to act with an additional minus sign on $A_B$ (again, just like in M-theory, where $C_3$ gets an additional $-$ sign for similar reasons \cite{Witten:1996md}). The $A_X$ field survives directly in 4d, and the BF coupling \eq{bc0} can be dualized to give a Stuckelberg term
\begin{equation}  \int d^4x  \frac{1}{R\, g_5^2} \vert d\phi_B-pA_X\vert^2\quad m_{A_X}^2=p^2 \frac{g_X^2}{g_5^2R}.\label{stuckel}\end{equation}
Thus, the 4d gauge field $A_X$ gets a mass. The couplings \eq{chb} reduce to couplings of the form
\begin{equation} \int\frac{\mathcal{A}_1}{6} F_X\wedge F_X \wedge \phi_B -\frac{\mathcal{A}_2}{24}p_1(R)\wedge \phi_B,\label{chb2}\end{equation}
so we see that $U(1)_X$ has become massive precisely because of the 4d Green-Schwarz mechanism.
 
It is important to note that, for the above mechanism to work, the anomaly polynomial of $U(1)_X$ must be severely constrained. It must be of the form \eq{chb}, i.e. it can have pure $U(1)_X^3$ and mixed $U(1)_X$-gravitational anomalies, but it \emph{cannot have any mixed anomalies with SM gauge fields}. This is similar to how the Horava-Witten factorization only works with very special groups \cite{Horava:1996ma} where all anomalies can be written purely in terms of $c_2$. In other words, this mechanism explains why B-L is a good approximate symmetry of the SM, and why it has no mixed anomalies with it, but it does have pure and mixed gravitational ones -- these are precisely the ones that can be cancelled by the Green-Schwarz mechanism. One would have expected that a random global symmetry of the SM would have  had mixed anomalies with SM gauge fields. The fact that B-L does not suggests it has an ulterior meaning, and the above mechanism is a scenario where this can be explained naturally without having any new particles. In this sense, it is similar to the mechanisms in \cite{Wang:2020mra,Wang:2025oow}.  Notice that, in this case, bulk massless particles charged under $U(1)_B$ cannot alter the anomaly, since their zero modes are not charged under the gauge fields.
 
The fact that B-L is massive is appealing, since it means that it is less constrained. Using \eq{stuckel} with $p=1$ together with $M_5R\sim10^{20}$ in Dark Dimension we obtain
\begin{equation} m_{A_X}\sim \frac{g_X}{\sqrt{\alpha}}\, \text{1 GeV}.\end{equation}
On the other hand, $g_X\sim 0.1$ since it unifies with the SM couplings. At this coupling, the lower bound on the mass comes from LHC, and is around 10 TeV. In turn, this leads to a bound in the 5d gauge coupling, %phrasing
\begin{equation} \sqrt{\alpha} \lesssim 10^{-5}.\label{bound5}\end{equation}
Notice that this is the value of the 5d gauge coupling in 5d Planck units, and that this does not unify with the SM gauge couplings unlike $g_X$. This is why it can take any value.
 
So far, the scenario seems quite appealing. However, there is a problem because we need to generate operators that break the B-L symmetry to give neutrino masses. For instance, we would like to generate the dimension 5 Weinberg operator,
\begin{equation} \mathcal{O}_5\sim \frac{1}{\Lambda_5} H^\dagger H L^\dagger L,\label{wein}\end{equation}
which would generate Majorana neutrino masses,  and determine the scale $\Lambda_5$. If this was a standard Higgsing scenario, where B-L is broken by the vev of some field, one would expect
\begin{equation}\Lambda_5^{-1}\sim \frac{f}{\Lambda^2_{UV}}\end{equation}
and $f$ is the symmetry breaking vev. This is what happens in a GUT with right-handed neutrinos, and indeed the effective operator obtained from integrating out the heavy right-handed neutrinos follows this scaling. However, based on string theory examples, whenever one has a 4d Green-Schwarz, the symmetry is usually broken by non-perturbative effects suppressed by exponential factors. To see this, consider a symmetry breaking operator, like \eq{wein}. Since the operator transforms with some phase under the spontaneously broken $B-L$ symmetry, to restore gauge invariance we must insert an exponential of the axion $\phi_B$,
\begin{equation} e^{i\phi_B\cdot q} H^\dagger H L^\dagger L,\label{wein2}\end{equation}
where $q$ is the $X$- charge of the operator being inserted. In a Higgsing mechanism, the $\phi_B$ is the phase of the Higgs field, so naturally it becomes an insertion of said field,
\begin{equation} e^{i\phi_B\cdot q}\,\rightarrow \Phi^q= f^q e^{i\phi_B\cdot q},\end{equation}
which justifies our earlier choice. However, in extra-dimensional models like the above, $e^{i\phi_B\cdot q}$ is typically coming from an instanton insertion,
\begin{equation} e^{i\phi_B\cdot q}\,\rightarrow \mathcal{P}\, e^{-S}e^{i\phi_B\cdot q},\end{equation}
where $S$ is the action of an instanton coupled to $\phi_B$, and $\mathcal{P}$ is some instanton prefactor with units of energy. In our case, the natural candidate would be an Euclidean  5d particle electrically charged under $A_B$ which is wrapping the interval, with action
\begin{equation} S\sim m_5\, R,\end{equation}
 If we take the value of $m_5$ suggested by Weak Gravity, $m_5\sim g_5\, M_5^{3/2}$, we have an instanton action $S\sim 10^{15}$, and neutrino masses unacceptably small.  One way out would be to take $m_5\sim 0$, so there would be a massless or light particle charged under $A_B$. This masslessness is quite unnatural, as it is not explained by any symmetry principle (even if the particle is  a fermion;  unlike in 4d, there is no chirality in five dimensions and the smallest fermion representation admits a $U(1)_B$-invariant Dirac mass). We could also lower the WGC bound by having the bulk gauge coupling $\alpha$ to be very small, much smaller than the bound of $10^{-10}$ in \eq{bound5}. This, too, would amount to a significant additional fine-tuning. Therefore, we have run into a corner; if we remove right-handed neutrinos in the Dark Dimension, and take anomaly cancellation of B-L seriously, we are forced to have a Green-Schwarz mechanism and in turn to introduce a strong fine-tuning in $m_5$ to explain neutrino masses.
 
If the Green-Schwarz mechanism model exposed here was the only alternative for a bulk neutrino, we would be forced back to the scenarios explained in Section \ref{s2} -- and, in particular, to the GUT scenario, which has none of the problems of the bulk right-handed neutrinos that motivated us to look for inflow solutions to the B-L problem. However, as explained at the beginning of this section, there is a natural alternative to this case, where the bulk gauge field transforms as an ordinary vector under parity; we will explore this possibility next.
 
 \subsection{The vector case}\label{good}
 
We will now assume that the bulk gauge field $A_B$ has Dirichlet boundary conditions for the longitudinal component at a boundary, as would be the case for a vector field transforming ordinarily under parity. In this case, dimensional reduction on an interval $S^1/\mathbb{Z}_2$ produces a four-dimensional vector (unlike in the case of the previous Subsection, where it produced an axion). This vector may be directly identified with the $U(1)_{B-L}$ gauge field. Therefore, in this scenario, the B-L gauge field lives in the bulk.
 
The first thing we should explain is anomaly cancellation: After dimensional reduction on the interval, the B-L gauge field becomes a 4d gauge field, and the SM fields live in an anomalous representation of it. It is also not possible to cancel the anomaly via 5d Chern-Simons terms yielding a 4d Green-Schwarz mechanism, as in the previous Subsection, since the parity transformation of the gauge field $A_B$ forbids their existence. Anomaly cancellation is far simpler in this case: Consider a 5d bulk fermion $\psi$, carrying charge $q$ under $U(1)_{B-L}$. Upon dimensional reduction on a circle, it yields a four-dimensional Dirac fermion, with both Weyl components carrying the same charge $q$. To reduce on $S^1/\mathbb{Z}_2$, we simply project onto the 4d circle modes invariant under parity, a transformation that acts as\footnote{In general, one can also choose a parity transformation that squares to fermion parity; that is known as a Pin$^-$ structure. A parity transformation acting as in the main text is a Pin$^+$ structure. An interval $S^1/\mathbb{Z}_2$ admits a Pin$^+$ but not a Pin$^-$ structure, since acting with the transformation twice it squares to $+1$ for Pin$^+$ and to $-1$ on fermions for Pin$^-$.}
\begin{equation}\psi(x)\,\rightarrow \pm\Gamma^5\psi(-x),\quad (\Gamma^5)^2=+1.\end{equation}
where the choice $\pm$ depends on the intrinsic parity of the 5d fermion. Therefore, one of the two chiralities is projected out, and we end up with a single four-dimensional Weyl fermion of charge $q$\footnote{An $U(1)_B$-invariant 5d mass term for the fermion is forbidden by parity as we will see below, which is why the massless four-dimensional zero mode is robust}. This is anomalous, so a suitable collection of 5d fermions can cancel the anomaly of the SM gauge fields. The simplest choice is to have three massless bulk fermions, each of charge one, which upon dimensional reduction yield three four-dimensional right-handed neutrinos, ensuring anomaly cancellation of $B-L$. Therefore, in this scenario, the presence of three massless bulk neutrinos is naturally explained. Anomaly cancellation between bulk fields and localized sectors (twisted sectors in orbifolds) is a familiar phenomenon in string theory, see e.g. \cite{Dixon:1985jw} for an early realization.

%The point: Mass terms that seem the same wrt Lorentz alone might not be the same wrt bigger groups. In 4d, a Majorana mass term psi^c psi can be turned into the ordinary, dirac one by rewriting psi^c =
 
The scenario described above also allows for the B-L symmetry to be spontaneously broken in the bulk, which is what is needed to generate neutrino masses. We will not commit to a specific mechanism for this breaking; we simply assume that it takes place at some scale, which we will refer to as the B-L breaking scale. As a result, the bulk gauge field $A_B$ acquires a mass $m_{B-L}$. For the natural value $\alpha\sim 1$ of the 5d gauge coupling \eq{norsd}, $m_{B-L}$ is of the order of this scale. Since the $U(1)$ symmetry is broken, there can be effective mass terms for the 5d neutrinos, in principle. The standard $U(1)$ invariant 5d Dirac mass term $\bar{\psi}\psi$ is allowed by the $U(1)$ charge, but not by the  parity symmetry, under which it transforms with a sign. On the other hand, the Dirac representation in five dimensions is isomorphic to its conjugate, so we can construct a mass term $\bar{\psi^c}\psi$. This mass term is allowed by parity, since the conjugate spinor $\psi^c$ transforms under parity with opposite sign to $\psi$, but it breaks the $U(1)_{B-L}$.  As a result, the mass term is also forbidden\footnote{Absence of fermion masses may suggest the presence of some anomaly. Since the Pin$^+$ bordism groups of a point are all torsion, this anomaly can only be torsion, and therefore it may be trivialized by a rescaling of the $B-L$ charge. Therefore the anomaly can be ignored; we compute it anyway in Appendix \ref{appa}.}.
Once B-L is broken, however, a Majorana mass term for the bulk neutrinos can be generated, with a mass $m_\psi$ which is expected to be of the order of the B-L breaking scale. We will remain agnostic about the B-L breaking scale and about the precise value of $m_\psi$, since, as we now show, the active neutrino masses are insensitive to it.
To obtain the active neutrino masses one must take into account the full tower of bulk modes coupling to the brane, rather than a single sterile state as in \eq{matrix}. Since this mass term is Lorentz invariant in five dimensions, the $n$-th KK level of $\psi$ consists of a pair of Weyl fermions with Dirac mass $n\, m_{KK}$ and Majorana mass $m_\psi$, and hence physical mass $\sqrt{m_\psi^2+n^2m_{KK}^2}$. Only the parity-even Weyl fermion of each pair (the one with the same B-L charge as the zero mode) couples to $\nu_L$ on the brane, with the Dirac mass $m_D= y\langle H\rangle/\sqrt{M_5 R}$ appearing in \eq{matrix}. For $m_\psi\gg m_D$, integrating out the $n$-th level contributes $m_D^2\, m_\psi/(m_\psi^2+n^2 m_{KK}^2)$ to the Majorana mass of the active neutrino, all levels contributing with the same sign. Summing over the tower one obtains a remarkably simple expression for the neutrino mass \cite{deGiorgi:2026isr}:
 \begin{equation} m_\nu\simeq \frac{y^2\langle H\rangle^2}{M_5 R}\sum_{n}\frac{m_\psi}{m_\psi^2+n^2m_{KK}^2}\approx \frac{y^2\langle H\rangle^2}{M_5}\,\coth\left(m_\psi R\right), \label{ewe4}\end{equation}
up to $\mathcal{O}(1)$ factors that depend on the position of the brane along the interval\footnote{For the SM brane at the end of an interval of length $R$, so that $m_{KK}=\pi/R$ and the sum runs over $n\in\mathbb{Z}$, the last expression is exact.}. The dependence on the sterile neutrino mass drops out\footnote{In the original version of this paper, the effect of integrating out massive modes were not considered, leading to incorrect conclusions, which we have corrected in the current version of the paper.  This was pointed out in \cite{deGiorgi:2026isr} and private communication with L. Anchordoqui and C. Gonzalez-Garcia, to whom we are grateful.} as soon as $m_\psi\gtrsim m_{KK}$: the neutrino mass saturates at the value $m_\nu\approx y^2\langle H^2\rangle/M_5$.  This is easy to understand from the five-dimensional point of view: the active neutrino mass is generated by the lepton-number violating part of the bulk fermion propagator between coincident points on the brane, $\int \frac{dp_5}{2\pi} \frac{m_\psi}{p_5^2+m_\psi^2}=\frac12$, which is independent of $m_\psi$, so that the neutrino mass is a local effect on the brane. 
 
Thus, the breaking of B-L leaves the prediction of \cite{Montero:2022prj} for the active neutrino masses unchanged: with $M_5\sim 10^{9-10}$ GeV, the observed neutrino masses are obtained for $y\sim 10^{-3}$, a value comparable to other Yukawa couplings of the SM, and the coincidence $m_\nu\sim m_{KK}\sim \Lambda^{1/4}$ can be understood as discussed in Section \ref{s2} and in \cite{Gonzalo:2022jac}. The main difference with the massless bulk fermion case is that the sterile neutrinos are now heavy, with masses set by the B-L breaking scale.
 
%Of course, the vacuum energy coming from $V$ is much too large, and has to be fine-tuned away. This is the standard cosmological constant problem, which the Dark Dimension scenario does not address. At any rate, we see how reasonable interactions can yield a natural explanation for neutrino masses based on a bulk $B-L$, while also having the appealing feature of tying together the vacuum energy, and the symmetry breaking of electroweak and $B-L$ symmetries.
 
We still need to compare this scenario with experimental bounds for a B-L gauge boson.  The four-dimensional gauge coupling is given by
\begin{equation}  g_{B-L}^2= \frac{\alpha}{M_5\, R}= \frac{\alpha}{10^{20}}.\end{equation}
The natural value of $\alpha\sim 1$ leads to a gauge coupling $g_{B-L}\sim 10^{-10}$, independently of the B-L breaking scale, while the mass of the $B-L$ gauge boson is of the order of the B-L breaking scale. As an example, if B-L is broken at the weak scale, we get a mass of order the Higgs scale for the $B-L$ gauge boson, $m_{B-L}\sim 100$ GeV. We stress that this is just an example: as we will see, a wide range of values of the B-L breaking scale is consistent with experiments.
 
\begin{figure}[!htb]
    \centering
     \includegraphics[width=0.9\textwidth]{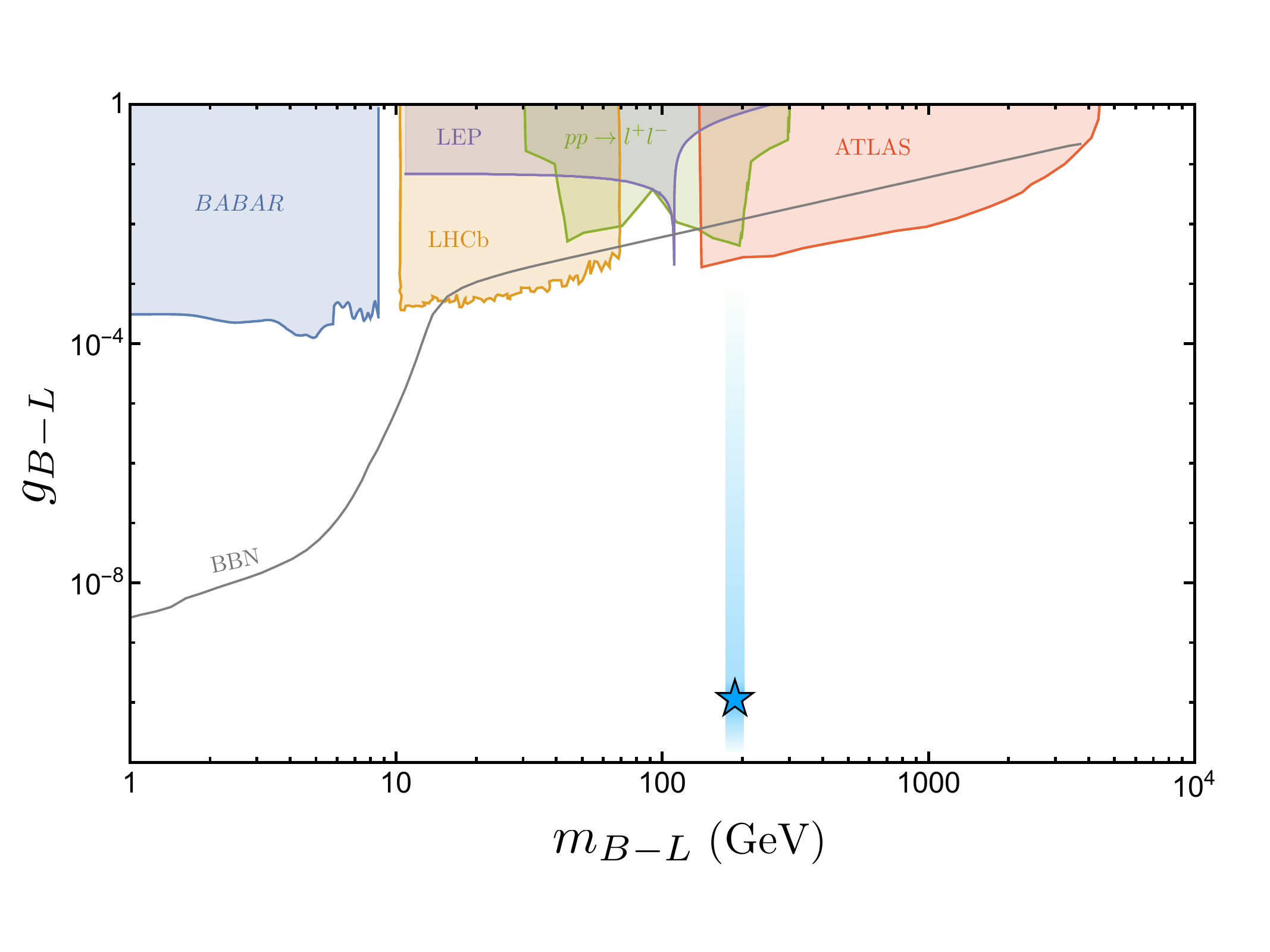}
     \vspace{-0.7cm}
    \caption{Theoretical prediction versus experimental constraints for the gauge coupling and mass of the $B-L$ gauge boson.
    We show a benchmark point at the natural value for the gauge coupling $g_{B-L}\sim 10^{-10}$ and a value $m_{B-L}\approx 100 GeV$, although the latter can be changed freely.
    The shaded band indicates the uncertainty associated with the precise value of $\alpha$ and the energy at which the gauge coupling is evaluated. We also show experimental limits from BABAR \cite{BaBar:2017tiz}, LHCb \cite{LHCb:2017trq}, Drell-Yan process in ATLAS \cite{Escudero:2018fwn,ATLAS:2017rue} ATLAS  \cite{ATLAS:2017fih}, LEP (taken from \cite{Asai:2023mzl,Wang:2024gvt}), and BBN limits taken from \cite{Heeck:2014zfa,Herbermann:2025uqz}.}
    \label{fig-BL}
\end{figure}

Constraints on the $(m_{B-L}, g_{B-L})$ parameter space can be found e.g. in \cite{Heeck:2014zfa,Ilten:2018crw,Escudero:2018fwn,Herbermann:2025uqz,Asai:2023mzl,Wang:2024gvt} and depicted in Figure \ref{fig-BL}. For mases around 100 GeV, the dominant constraints come from LHC searches.  The value $g_{B-L}\sim10^{-10}$ predicted by the Dark Dimension is safely within the allowed region for our example $m_{B-L}\approx 100$ GeV, and indeed for any value of $m_{B-L}$ in the range shown in the figure.
One might worry that, since $m_{B-L}\sim\, 100$  GeV in our example, future LHC searches (which see up to  $\approx100$ times this energy) may also see the higher KK modes of the $B-L$ vector boson, thereby dramatically increasing the production cross-section.  If one has $N$ almost degenerate copies of the vector  participating, a tree-level cross-section scales like
\begin{equation}\sigma_N\sim N^2\, \sigma_1,\end{equation}
since the Feynman diagrams are added coherently. The number of KK modes contributing at energy $s$ is
\begin{equation} (N-1)^2\sim \frac{s-m^2_{B-L}}{m_{KK}^2},\end{equation}
where the $-1$ shift on $N$ accounts for the zero mode. This translates to an effective running of the gauge coupling,
\begin{equation}g_{B-L,\text{effective}}^{2}(s)=\frac{g_{B-L}^2}{m_{KK}} \left(m_{KK}+\sqrt{s-m^2_{B-L}}\right)=\alpha\,\frac{m_{KK}+\sqrt{s-m^2_{B-L}}}{M_5} .\end{equation}
At $s\gg m_{B-L}^2$, we recognize the linear running of $g_{B-L}^2$ with energy characteristic of 5d gauge theory. Since $m_{KK}$ is so small, even at energies $\sqrt{s}$ slightly above $m_{B-L}$ we will get a significant enhancement of the effective gauge coupling. For $\alpha\sim1$, setting $\sqrt{s}\sim 10$ TeV leads to $g_{B-L,\text{effective}}\sim10^{-3}$. While this is indeed much larger than our naive estimate, it is still within the allowed bounds, but perhaps within experimental reach of future colliders. At any rate, the analysis we just carried out should be regarded as a first step only, since the precise effects of the KK towers of both B-L and right-handed bulk neutrino fields will depend on the details of the process under consideration (for instance, for LEP bounds, one should use the LEP $\sqrt{s}\sim 100$ GeV, for which the enhancement is at most $g_{B-L,\text{effective}}\sim 10^{-4}$, again below the experimental bounds). Astrophysical bounds on B-L are not relevant as long as the B-L gauge boson is too heavy to be produced in supernovae \cite{Esseili:2023ldf}, and neither are BBN bounds for the same reason; this is the case for $m_{B-L}$ above the GeV scale, as in our example.   
 
 \section{Concluding thoughts on neutrino masses and GUTs}\label{s4}
 In the context of the Dark Dimension scenario, it is natural to ask if the existence of the mesoscopic fifth dimension can naturally accommodate both the B-L gauge symmetry and provide a new mechanism for neutrino masses. As we have seen in this paper, there are several ways to do this -- a standard brane GUT scenario, a bulk B-L gauge field made massive by a Green-Schwarz mechanism, and a bulk $U(1)$ made massive by Higgsing. A brane GUT where B-L appears as part of the gauge symmetry (such as $SO(10)$) and right-handed neutrinos as part of the GUT multiplet with the mass of the neutrinos being generated via the usual seesaw mechanism is completely viable, and it also explains coincidence of the number of right-handed neutrinos with the number of generations. It can connect to specifics of the dark dimension scenario if we use the fact that the $M_5$ is the geometric means of the GUT scale and weak scale as in \cite{Heckman:2024trz} leading to the standard GUT seesaw explanation of neutrino masses. In contrast, the Green-Schwarz case provides an explanation for B-L anomaly cancelation but  leads to neutrino masses that are way too small.
 
The B-L Higgsing scenario is also appealing, due to the following:
\begin{itemize}
\item It explains why there are three right-handed bulk neutrinos, via anomaly cancellation with the SM fields;
\item The natural value of the B-L gauge coupling is $g_{B-L}\sim10^{-10}$, independently of the B-L breaking scale, which sets the mass of the B-L gauge boson and of the bulk sterile neutrinos. For instance, B-L breaking at the weak scale gives $m_{B-L}\sim m_{\text{Higgs}}$; these values are not excluded by experiments, and neither is a wide range of other values of the B-L breaking scale.  
\item Once the full tower of bulk modes is taken into account, the active neutrino masses are insensitive to the mass of the bulk sterile neutrinos and are given by the original Dark Dimension estimate
\begin{equation}m_{\nu}\approx \frac{y^2\langle H\rangle^2}{M_5},\end{equation}
which reproduces the observed neutrino masses for $y\sim 10^{-3}$. Therefore, the breaking of B-L, which makes the sterile neutrinos heavy, does not spoil the relationship $m_\nu\sim m_{KK}\sim\Lambda^{1/4}$ of the original proposal, and makes the hierarchy between sterile and active neutrino masses required by the seesaw mechanism automatic.
 \end{itemize}
 
It is reassuring that these features come out in what is perhaps the simplest embedding of a B-L bulk gauge field with a single extra dimension, once the Dark Dimension scaling $m_{KK}\sim \Lambda^{1/4}$ is imposed. This simplicity, which has appeared in other aspects of the Dark Dimension scenario, suggests that it may be on the right track. In this line, it would be interesting to test this scenario in the context of recent searches for extra-dimensional sterile neutrinos \cite{Elacmaz:2025ihm,deGiorgi:2025xgp}\footnote{We believe some of the Majorana results of \cite{deGiorgi:2025xgp} and related references likely need revision, since the prediction in a simple dimensional reduction of a 4d KK mode much lighter than the 5d bulk fermion mass violates Lorentz invariance.}.
 
We finish with a couple comments. First, the insensitivity of the active neutrino masses to the bulk sterile neutrino mass is special to a \emph{single} large extra dimension: with two large extra dimensions, the lepton-number violating part of the bulk fermion propagator at coincident points on the brane is proportional to $m_\psi$ (up to a logarithm of the cutoff), so that the neutrino masses would depend on the B-L breaking scale and the simple estimate $m_\nu\sim y^2\langle H\rangle^2/M_5$ would be lost.
 
Another thought is that, although not strictly required by the model, it is natural to put the SM brane at the end of an interval (see e.g. \cite{Schwarz:2024tet}). In this case, there will be another boundary and, consequently, another brane at the far side of the Dark Dimension. This may or may not have gapless degrees of freedom (think of e.g. the T-dual of type I string theory on a circle), but its degrees of freedom may couple as B-L in a similar way to the SM brane. One should keep in mind this possibility as the B-L bulk scenario is explored further.

\subsubsection*{Acknowledgments}

We are grateful to Miguel Escudero and Luis Anchordoqui for enlightening discussions and to Maksym Ovchynnikov for valuable help in relation to the figure.
The authors thank the Simons Center for Geometry and Physics for hospitality
during the Summer Workshop ’25. MM and IV also thank the Harvard Swampland Initiative for hospitality during the early stages of this work.
 
The work of CV is supported in part by a grant from the Simons Foundation (602883,CV) and a gift from the DellaPietra Foundation. MM and IV
thank the Spanish Research Agency (Agencia Estatal de Investigacion) through
the grants IFT Centro de Excelencia Severo Ochoa CEX2020-001007-S, PID2021-123017NB-I00 and PID2024-156043NB-I00, funded by MCIN/AEI/10.13039/501100011033 and by ERDF A way of
making Europe.  MM is currently supported by
the RyC grant RYC2022-037545-I and project EUR2024-153547 from the AEI. The work of I.V. was supported by the ERC Starting Grant QGuide101042568 - StG 2021, and the Project
ATR2023-145703 funded by MCIN /AEI /10.13039/501100011033.
 
\appendix
 
\section{Anomalies for the bulk theory}\label{appa}
\begin{figure}[htb!]
    \centering
     \includegraphics[width=0.7\textwidth]{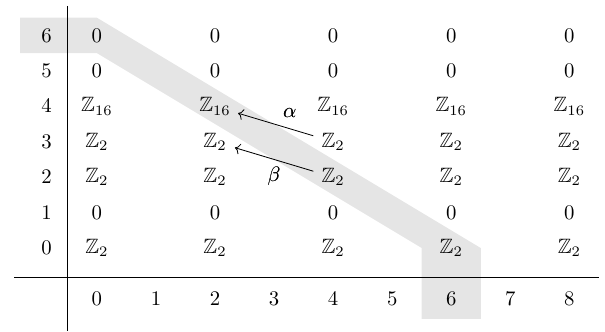}
    \caption{$E_2$ page of the spectral sequence used to compute the bordism group controlling bulk anomalies.}
    \label{fig1-ap}
\end{figure}

In Section \ref{good}, we introduced a five-dimensional bulk theory with Pin$^+$ reflections coupled to $U(1)$ gauge fields. To determine if there are any anomaly cancellation constraints on this fermion spectrum we must compute the bordism group $\Omega_6^{\text{Pin}^+}(BU(1))$. This can be done via the Atiyah-Hirzebruch spectral sequence (see e.g. \cite{Garcia-Etxebarria:2018ajm}). The second page of this sequence is displayed in Figure \ref{fig1-ap}.
At degree six, there are three possible entries, and just two potentially relevant differentials. The differential $\beta$ has as target the only class in dimension 5, but this class (corresponding to $KB\times S^1\times S^2$ with one unit of $U(1)$ flux along the $S^2$) vanishes: just take the $S^1\times S^2$ factor and view it as a class in $\Omega_3^{\text{Spin}}(BU(1))=0$. This means the differential is nontrivial, and the source class in dimension 6 is also killed. The source of the differential $\alpha$ is $KB\times S^2 \times S^2\times S^1$, and is trivial for a similar reason. On the other hand, the class in $(6,2)$, which could be the source of a differential to the source of $\alpha$, is represented by $\mathbb{CP}^3\times KB$ with the canonical $U(1)$ bundle on the $\mathbb{CP}^3$ factor, which is non-trivial since it is detected by the 8d $\eta$ invariant of a Dirac fermion of charge 2 (so that the Dirac index is 1). As a result, $\alpha$ is also non-trivial, and we are looking at an extension of a single $\mathbb{Z}_2$ factor by $\mathbb{Z}_{8}$. The $\mathbb{Z}_2$ is represented by $\mathbb{CP}^3$ with the canonical $U(1)$ bundle on top,  and detected by the bordism invariant $\int c_1^3\,\text{mod}\,2$. So
\begin{equation}\Omega_6^{\text{Pin}^+}(BU(1))=\mathbb{Z}_{8}\oplus\mathbb{Z}_2.\end{equation}
The anomaly cancellation condition for a set of 5d fermions with charges $q_i$ is that $\sum_i q_i\equiv 0\,\text{mod}\, 8$. As emphasized in the main text, this can be trivialized, even for a single fermion, if we declare its charge is eight times the fundamental charge, so it poses no constraints on the spectrum (equivalently, we are declaring that the smallest possible B-L charge in the bulk is $1/8$ of the charges of SM fields).

\bibliographystyle{utphys}
\bibliography{BLrefs.bib}
 
\end{document}